# Electromotive forces generated in *3d*-transition ferromagnetic metal films themselves under their ferromagnetic resonance

Kazunari Kanagawa[1], Yoshio Teki[2], and Eiji Shikoh[1,a)]

[1]*Graduate School of Engineering, Osaka City University, 3-3-138 Sugimoto, Sumiyoshi-ku, Osaka 558-8585, Japan*

[2]*Graduate School of Science, Osaka City University, 3-3-138 Sugimoto, Sumiyoshi-ku, Osaka 558-8585, Japan*

We report the electromotive force (EMF) properties generated in *3d*-transition ferromagnetic metal (FM = Fe, Co, and $Ni_{80}Fe_{20}$) films themselves under their ferromagnetic resonance (FMR). For Fe and Co films, the EMF due to the anomalous-Hall effect is dominantly generated under their FMR. Meanwhile, for a $Ni_{80}Fe_{20}$ film, the EMF due to the inverse spin-Hall effect in the $Ni_{80}Fe_{20}$ film itself under the FMR is mainly generated. This tendency is qualitatively explained with differences of the spin polarization, the spin Hall conductivity, the anomalous Hall conductivity, the magnetization saturation, and the resistivity of the FM films.

a) E-mail: shikoh@elec.eng.osaka-cu.ac.jp

In spintronics, recently, the spin-pumping with the ferromagnetic resonance (FMR) and the inverse spin-Hall effect (ISHE) have become powerful techniques to generate a spin current and to detect the spin current, respectively.[1-12] First, those techniques were used for spin injection from a ferromagnetic metal $Ni_{81}Fe_{19}$ as a spin injector into a nonmagnetic Pt layer working as a spin detector, by using the $Ni_{81}Fe_{19}$/Pt bi-layer structure samples.[1] In the spin injection by using the spin-pumping with the FMR, the conductance mismatch problem between the ferromagnetic material and the target-material, which causes lowering the spin injection efficiency in the case of an electrical spin injection method,[13,14] is negligible.[4,5,7,8,10,12] Therefore, while those techniques are very fundamental, those apply to investigate the spin transport property of various materials, for example, by using the ferromagnetic metal (FM) /target-material/non-magnetic metal (NM) junction structure samples.[5-7,12] Meanwhile, it was discovered that the electromotive force (EMF) was generated in a single layer $Ni_{80}Fe_{20}$ film itself under the FMR.[15] The suggested EMF generation mechanism in the single-layer $Ni_{80}Fe_{20}$ film itself under the FMR is as follows; a spin current is generated due to the magnetic inhomogeneity of the $Ni_{80}Fe_{20}$ film under the FMR condition and converted to a charge current with the spin-orbit interaction of $Ni_{80}Fe_{20}$ film, that is, the ISHE in the $Ni_{80}Fe_{20}$ film.[15] Previously, by using the $Y_3Fe_5O_{12}$(YIG)/ferromagnet "bi-layer" structure samples, the ISHE generated in 3d-transition ferromagnetic metal films of Fe, Co, $Ni_{80}Fe_{20}$ and Ni under the FMR of the YIG was observed,[9,11] while the EMF generated in a single-layer ferromagnetic metal film itself under the FMR is not investigated except for the $Ni_{80}Fe_{20}$ film.[15] If those EMF generation phenomena by using the spin-pump with the FMR of a FM material were applied for practical use, the single-layer structure is better than any other multi-layer structures, in terms of the material saving and easiness of manufacturing devices. In this study, the EMF properties generated in single-layer 3d-transition FM films themselves under the respective FMR were investigated.

Our sample structure and experimental set up are illustrated in Figure 1. On a thermally-oxidized silicon substrate, a ferromagnetic metal (FM = Fe, Co, and $Ni_{80}Fe_{20}$) was deposited to a thickness of 25 nm by using a conventional DC magnetron sputtering system. No protection layer was formed on the FM films, as similar to the previous study.[15] After forming the FM films, the sample substrates were cut as a rectangular shape of $4.0 \times 1.5$ mm$^2$, to measure the physical properties.

A sample substrate was set into the microwave $TE_{011}$-mode cavity of an electron spin resonance (ESR) system (JEOL, JES-TE300) to excite the FMR of the sample. The microwave frequency $f$ to excite the FMR was 9.45 GHz. The EMF property of the FM sample was measured by using a nano-voltmeter (Keithley Instruments, 2182A). Leading wires to detect the output voltage properties from a sample were directly attached with silver paste at both ends of the film sample. All of the measurements were performed at RT.

Figures 2 (a)-(c) show the FMR spectra of samples at an external magnetic field orientation angle $\theta$ of 0°, with the microwave power $P_{mW}$ of 200 mW to excite the respective FMR. Figs. 2 (a), (b) and (c) are for an Fe sample, for a Co sample, and for a $Ni_{80}Fe_{20}$ sample, respectively. As expected, the FMR was observed in all FM films at the respective FMR field $H_{FMR}$ of 1061 Oe for the Fe, 1094 Oe for the Co and 472.8 Oe for the $Ni_{80}Fe_{20}$. The saturation magnetization $M_S$ was estimated to be 1061 emu/cc for Fe, 1094 emu/cc for Co and 472.8 emu/cc for $Ni_{80}Fe_{20}$ by using the following equation:[2,3]

$$\omega_0 = \gamma\sqrt{H_{FMR}(H_{FMR} + 4\pi M_S)}, \qquad (1)$$

where $\omega_0$ (= $2\pi f$) and $\gamma$ are respectively, the angular frequency of the microwave and the gyromagnetic ratio of the respective FMs.

Figs. 2 (d)-(f) show the output voltage properties at the $\theta$ of 0° and 180°, with the $P_{mW}$ of 200 mW for the excitation of the respective FMR. Figs. 2 (d), (e) and (f) are for an Fe sample, for a Co sample, and for a $Ni_{80}Fe_{20}$ sample, respectively. For the experimental data (open circles), components which do not relate to the magnetic field orientation angles are removed by using the eq.(2):

$$V = \frac{V_0 - V_{180}}{2}, \quad (2)$$

where the $V_0$ and $V_{180}$ correspond to the EMFs at the $\theta$ of 0° and 180°, respectively. Output voltages were observed in all FM films themselves under the respective FMR. The polarity of output voltages is inverted in all FM films against the magnetization reversal of the respective FM films. The output voltages increased with the increase of $P_{mW}$. These polarity inversion to the magnetization reversal and $P_{mW}$ dependence of output voltages are similar to previous studies using the spin-pump and ISHE.[3-8,10,12,15] To analyze those output voltage properties, the data in Figs. 2 (d)-(f) were fitted by the following equation:

$$V(H) = V_{ISHE}\frac{\Gamma^2}{(H-H_{FMR})^2+\Gamma^2} + V_{AHE}\frac{-2\Gamma(H-H_{FMR})}{(H-H_{FMR})^2+\Gamma^2} + V_{BG}, \quad (3)$$

where the first and second terms of the eq. (3) correspond to the EMF due to the ISHE and the EMF due to the extra ordinary Hall effect in FMs, that is, the anomalous Hall effect (AHE) in FMs, respectively.[1,4-8,10,12,15] The $V_{ISHE}$ and $V_{AHE}$ indicate the magnitude of the EMF due to the ISHE and that due to the AHE, respectively. $\Gamma$ is a damping constant in these fittings. The ISHE term is a Lorenz function, in other words, symmetry to the $H_{FMR}$, while the AHE term is derivative of a Lorenz function, which is anti-symmetry to the $H_{FMR}$. $V_{BG}$ is background signals on experiments, which are independent of the external magnetic field. The fitting results are drawn with the solid lines in Figs. 2 (d)-(f). The above analysis indicated that the electromotive forces generated in the 3d-transition FM metal films under their FMR were successfully observed. The ISHE properties of the FM films were similar to the studies by using "bi-layer" structure samples.[9,11]

Table 1 shows the summary of the analysis, where the $V_{ISHE}$, $V_{AHE}$, the absolute value of the ratio of $|V_{ISHE}/V_{AHE}|$, $M_S$, the ratio to the $V_{ISHE}$ of Co, and the ratio to the $V_{AHE}$ of $Ni_{80}Fe_{20}$ are described. From the values of the $|V_{ISHE}/V_{AHE}|$, we can say that the AHE is dominant for the Fe and Co samples, while the ISHE is dominant for the $Ni_{80}Fe_{20}$ sample. This tendency has the reproducibility and may be due to the difference of the spin polarization ($P_S$), spin-Hall conductivity ($\sigma_{SHE}$), anomalous Hall conductivity ($\sigma_{AHE}$), $M_S$, and resistivity ($\rho$) of the FM films. Table 2 shows the data of the $P_S$,[16,17] the $\sigma_{SHE}$,[15,18] the $\sigma_{AHE}$,[15,18] and the $\rho$, where only the $\rho$ values were experimentally obtained in this study.

In the ISHE regime,[1] the spin current density $j_S^0$ is converted to a charge current density $j_C$, as follows:[3,5]

$$\vec{j_C} \propto \theta_{SHE} \vec{j_S^0} \times \vec{\sigma}, \quad (4)$$

where the $\vec{\sigma}$ and $\theta_{SHE}$ are the spin polarization vector of the spin current and the spin-Hall angle, which is a kind of a conversion efficiency from the $j_S^0$ to the $j_C$. Thus, the absolute value of the $j_C$ can be expressed as follows:

$$|j_C| \propto \theta_{SHE} j_s^0. \quad (5)$$

The $\theta_{SHE}$ is equal to $\dfrac{\sigma_{SHE}}{\sigma_C}$, where the $\sigma_C$ is the electrical conductivity of a FM film and corresponds to $1/\rho$. Thus, the eq. (5) is rewritten as follows;

$$\left|j_C\right| \propto \rho \sigma_{SHE} j_s^0. \qquad (6)$$

In this study, the ISHE is not observed as a charge current but as an EMF via the sample resistance. Under an assumption that the $j_S^0$ is approximately proportional to the $P_S$ in a FM film, the absolute value of $V_{ISHE}$ is expressed as follows;

$$\left|V_{ISHE}\right| \propto \frac{l\rho^2 P_S \sigma_{SHE}}{S}, \qquad (7)$$

where the $l$ is the length of an FM sample (4 mm in this study), and the $S$ is the sectional area of the FM sample (1.5 mm × 25 nm). Using the values in the Table 2, the $V_{ISHE}$ values of Fe, Co, and $Ni_{80}Fe_{20}$ samples to the $V_{ISHE}$ for a Co sample ratio is estimated to be Fe : Co : $Ni_{80}Fe_{20}$ = 8 : 1 : 6, while the experimentally obtained data were Fe : Co : $Ni_{80}Fe_{20}$ = 36 : 1 : 13 as shown in the Table 1. Those were qualitatively consistent, although the quantitative consistency lacks.

The $V_{AHE}$ in FM films is simply described as follows:[19]

$$V_{AHE} \propto M_S I_C, \qquad (8)$$

where the $I_C$ is a charge current in an FM film. In this study, the $I_C$ is generated due to the ISHE in FMs. Therefore, the eq. (8) can be rewritten as follows;

$$V_{AHE} \propto \frac{M_S V_{ISHE}}{\rho}. \qquad (9)$$

Similarly to the $V_{ISHE}$, using the values in the Table 2 for parameters in the eq.(9), the $V_{AHE}$ values to the $V_{AHE}$ for a $Ni_{80}Fe_{20}$ sample ratio is estimated to be Fe : Co : $Ni_{80}Fe_{20}$ = 5 : 1.1 : 1, while the values ratio on the experiments is Fe : Co : $Ni_{80}Fe_{20}$ = 134 : 11 : 1 as shown in the Table 1. The tendency of the AHE among the Fe, Co, and $Ni_{80}Fe_{20}$ samples was also qualitatively consistent between the above consideration and our experiments. The discrepancies between the above estimation and experiments for both the $V_{ISHE}$ and $V_{AHE}$ may come from the difference of the magneto-crystalline anisotropy of the FM films. In general, it is negligible for $Ni_{80}Fe_{20}$, while it strongly affects to the magnetic properties for Fe and Co films. The $M_S$ values were estimated by using the eq. (1), which is established for a uniform magnetization rotation mode and not considered the strong correlation between the neighbor spins. The way how to take the magneto-crystalline anisotropy of FM films into account must be found. Also, the polarities of the respective EMFs have not considered in this study, yet. However, the polarities of the output voltages are different among the researches.[9,11] Because of the lack of amount of the related studies, the discussion about the polarity of output voltages is a next issue. For further investigation, other *3d*-transition FM films with different *3d*-electron numbers are tested.

Finally, we compared the $V_{ISHE}$ value of the $Ni_{80}Fe_{20}$ sample in this study with the $V_{ISHE}$ of some Ni-Fe alloy/NM multi-layer samples in previous studies.[1-3] In this study, the $V_{ISHE}$ was estimated to be 14.7 µV for a single layer $Ni_{80}Fe_{20}$ film sample, while the $V_{ISHE}$ was estimated to be 10 µV ~ 30 µV for "bi-layer" $Ni_{80}Fe_{20}$/Pt or $Ni_{80}Fe_{20}$/Pd samples.[1-3] Thus, it was indicated that the single FM layer structure under the FMR generates high enough EMF compared to the FM/NM multi-layer structure. That is, we successfully demonstrated a better method to obtain the EMF under the FMR than using any other multi-layer structures, which has a merit of the material saving, and easiness of manufacturing devices.

The EMF properties generated in Fe, Co, and $Ni_{80}Fe_{20}$ films themselves under their FMR were investigated. For Fe and Co films, the EMF due to the AHE was dominantly generated under their FMR, while for $Ni_{80}Fe_{20}$ films, the EMF due to the ISHE in the $Ni_{80}Fe_{20}$ film itself under the FMR was mainly generated. This tendency was qualitatively explained with the $P_S$, the $\sigma_{SHE}$, the $\sigma_{AHE}$, the $M_S$, and the $\rho$ of the FM films.

[Acknowledgement]

This research was partly supported by a Grant-in-Aid from the Japan Society for the Promotion of Science (JSPS) for Scientific Research (B) (26286039).

References

[1] E. Saitoh, M. Ueda, H. Miyajima, and G. Tatara, Appl. Phys. Lett. **88**, 182509 (2006).

[2] K. Ando, Y. Kajiwara, S. Takahashi, S. Maekawa, K. Takemoto, M. Takatsu, and E. Saitoh, Phys. Rev. B **78**, 014413 (2008).

[3] K. Ando, and E. Saitoh, J. Appl. Phys. **108**, 113925 (2010).

[4] K. Ando, and E. Saitoh, Nat. Commun. **3**, 629 (2012).

[5] E. Shikoh, K. Ando, K. Kubo, E. Saitoh, T. Shinjo, and M. Shiraishi, Phys. Rev. Lett. **110**, 127201 (2013).

[6] Y. Kitamura, E. Shikoh, Y. Ando, T. Shinjo, and M. Shiraishi, Sci. Rep. **3**, 1739 (2013)

[7] Z. Tang, E. Shikoh, H. Ago, K. Kawahara, Y. Ando, T. Shinjo, and M. Shiraishi, Phys. Rev. B **87**, 140101 (R) (2013).

[8] J.C. Rojas-Sánchez, M. Cubukcu, A. Jain, C. Vergnaud, C. Portemont, C. Ducruet, A. Marty, L. Vila, J.-P. Attané, E. Augendre, G. Desfonds, S. Gambarelli, G. Jaffrès, J.-M. George, and M. Jamet, Phys. Rev. B **88**, 064403 (2013).

[9] B.F. Miao, S.Y. Huang, D. Qu, and C.L. Chien, Phys. Rev. Lett. **111**, 066602 (2013).

[10] Y. Ando, K. Ichiba, S. Yamada, E. Shikoh, T. Shinjo, K. Hamaya, and M. Shiraishi, Phys. Rev. B **88**, 140406 (R) (2013).

[11] H. Wang, C. Du, P.C. Hammel, and F. Yang, Appl. Phys. Lett. **104**, 202405 (2014).

[12]Y. Tani, Y. Teki, and E. Shikoh, Appl. Phys. Lett. **107**, 242406 (2015).

[13]G. Schmidt, D. Ferrand, L.W. Molenkamp, A.T. Filip, and B.J. van Wees, Phys. Rev. B **62**, R4790 (2000).

[14]A. Fert, and H. Jaffrès, Phys. Rev. B **64**, 184420 (2001).

[15]A. Tsukahara, Y. Ando, Y. Kitamura, H. Emoto, E. Shikoh, M. P. Delmo, T. Shinjo, and M. Shiraishi, Phys. Rev. B **89**, 235317 (2014).

[16]P.M. Tedrow, and R. Meservey, Phys. Rev. B **7,** 318 (1973).

[17]E. Villamor, M. Isasa, L.E. Hueso, and F. Casanova, Phys. Rev. B **88**, 184411 (2013).

[18]T. Naito, D.S. Hirashima, and H. Kontani, Phys. Rev. B **78**, 014413 (2008).

[19]R. Karplus, and J. M. Luttinger, Phys. Rev. **95**, 1154 (1954).

Figure captions:

Fig. 1. (Color online) A schematic illustration of our single ferromagnetic metal (FM) layer sample and experimental set up. The dimensions of the FM (Fe, Co and $Ni_{80}Fe_{20}$) layer are 1.5 mm × 4.0 mm and the thickness is 25 nm. Two electrodes are attached on both ends of the FM film using silver paste.

Fig. 2. (Color online) (a)-(c) FMR spectra of (a) an Fe sample, (b) a Co sample, and (c) a $Ni_{80}Fe_{20}$ sample under the microwave power of 200 mW. The $I$ is the microwave absorption intensity. (d)-(f) Static magnetic field $H$ dependence of the electromotive force (EMF), $V$, for $\theta$ = 0° (red line and circles) and 180° (blue line and circles). (d), (e) and (f) are for an Fe sample, for a Co sample, and for a $Ni_{80}Fe_{20}$ sample, respectively.

Table 1. The analysis results for our Fe, Co, and $Ni_{80}Fe_{20}$ samples.

Table 2. The parameters of Fe, Co, and $Ni_{80}Fe_{20}$ samples. Only $\rho$ values are obtained in this study.

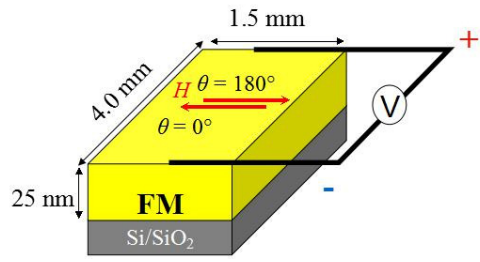

K. Kanagawa, et al.: FIG. 1.

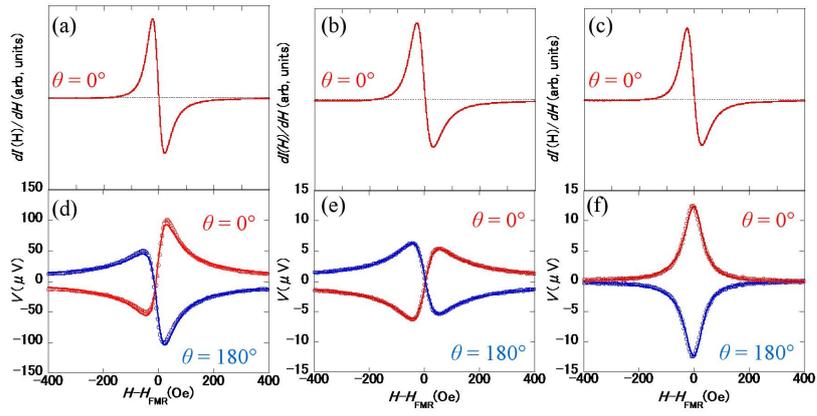

K. Kanagawa, et al.: FIG. 2.

|  | **Fe** | **Co** | **Ni$_{80}$Fe$_{20}$** |
|---|---|---|---|
| $V_{\text{ISHE}}$ (μV) | 34.0 | -0.939 | 12.4 |
| $V_{\text{AHE}}$ (μV) | -73.1 | -5.85 | 0.544 |
| The ratio to the $V_{\text{ISHE}}$ of Co | 36 | 1 | 13 |
| The ratio to the $V_{\text{AHE}}$ of Ni$_{80}$Fe$_{20}$ | 134 | 11 | 1 |
| $|V_{\text{ISHE}}/V_{\text{AHE}}|$ | 0.466 | 0.161 | 22.8 |
| $M_{\text{S}}$ (emu/cc) | 1061 | 1094 | 472.8 |

K. Kanagawa, et al.: Table 1.

|  | **Fe** | **Co** | **Ni$_{80}$Fe$_{20}$** |
|---|---:|---:|---:|
| $P_S$ [16,17] | 0.44 | 0.34 | 0.38 |
| $\sigma_{SHE}$ ($\Omega^{-1}$cm$^{-1}$) [15,18] | 400 | 200 | 133 |
| $\rho$ ($\Omega$cm) (experimental data) | 7.2×10$^{-5}$ | 4.2×10$^{-5}$ | 12×10$^{-5}$ |
| $\sigma_{AHE}$ ($\Omega^{-1}$cm$^{-1}$) [15,18] | 806 | 341 | 73 |

K. Kanagawa, et al.: Table 2.